\documentclass[english,aps,prb,superscriptaddress,twocolumn]{revtex4-1}
\usepackage[T1]{fontenc}
\usepackage[latin9]{inputenc}
\usepackage{prettyref}
\usepackage{amsmath}
\usepackage{amssymb}
\usepackage{graphicx}
\usepackage{esint}

\makeatletter

\newcommand{\lyxdot}{.}

 \@ifundefined{textcolor}{}
 {%
   \definecolor{BLACK}{gray}{0}
   \definecolor{WHITE}{gray}{1}
   \definecolor{RED}{rgb}{1,0,0}
   \definecolor{GREEN}{rgb}{0,1,0}
   \definecolor{BLUE}{rgb}{0,0,1}
   \definecolor{CYAN}{cmyk}{1,0,0,0}
   \definecolor{MAGENTA}{cmyk}{0,1,0,0}
   \definecolor{YELLOW}{cmyk}{0,0,1,0}
 }

\@ifundefined{definecolor}{\usepackage{color}}{}
\usepackage{babel}

\usepackage{babel}

\usepackage{babel}

\usepackage{babel}

\usepackage{babel}

\makeatother

\usepackage{babel}
\begin{document}

\preprint{This line only printed with preprint option}

\title{Green's functions from real-time bold-line Monte Carlo}

\author{Guy Cohen}

\affiliation{Department of Chemistry, Columbia University, New York, New York
10027, U.S.A.}

\affiliation{Department of Physics, Columbia University, New York, New York 10027,
U.S.A.}

\author{David R. Reichman}

\affiliation{Department of Chemistry, Columbia University, New York, New York
10027, U.S.A.}

\author{Andrew J. Millis}

\affiliation{Department of Physics, Columbia University, New York, New York 10027,
U.S.A.}

\author{Emanuel Gull}

\affiliation{Department of Physics, University of Michigan, Ann Arbor, MI 48109,
U.S.A.}
\begin{abstract}
We present two methods for computing two-time correlation functions
or Green's functions from real time bold-line continuous time quantum
Monte Carlo. One method is a formally exact generalized auxiliary
lead formalism by which spectral properties may be obtained from single-time
observables. The other involves the evaluation of diagrams contributing
to two-time observables directly on the Keldysh contour. Additionally,
we provide a detailed description of the bold-line Monte Carlo method.
Our methods are general and numerically exact, and able to reliably
resolve high-energy features such as band edges. We compare the spectral
functions obtained from real time methods to analytically continued
spectral functions obtained from imaginary time Monte Carlo, thus
probing the limits of analytic continuation.
\end{abstract}
\maketitle

\section{Introduction\label{sec:Introduction}}

Strongly correlated electron materials exhibit fascinating collective
behavior which has long challenged our understanding.\cite{dagotto_complexity_2005}
A few notable examples include Mott metal--insulator transitions in
transition metal oxides,\cite{imada_metal-insulator_1998} colossal
magnetoresistance in perovskite manganites,\cite{dagotto_nanoscale_2003}
quantum criticality in heavy fermion systems\cite{custers_break-up_2003}
and high temperature superconductivity in copper oxides\cite{bednorz_possible_1986}
and pnictides.\cite{kamihara_iron-based_2008} Of central experimental
interest are dynamical properties, both of excited states near the
Fermi energy and at highly excited energies. However, the theoretical
description of correlated electron materials has proven difficult:
because of the absence of a small parameter, perturbation theory is
in general unreliable. Traditional materials science techniques, among
them the density functional theory (DFT)\cite{hohenberg_inhomogeneous_1964,kohn_self-consistent_1965}
and the GW approximation,\cite{hedin_new_1965} do not capture strong
correlation effects, while standard numerically exact lattice methods\cite{blankenbecler_monte_1981,white_density_1992,corboz_simulation_2010}
are limited to small lattices, high symmetry points or one dimensional
systems.

An alternative to directly determining quantities in a correlated
quantum lattice model is provided by the dynamical mean field theory
(DMFT).\cite{metzner_correlated_1989,georges_hubbard_1992,georges_dynamical_1996}
In the approximation that correlations are local, the physics of the
lattice model can be mapped onto a numerically tractable impurity
model: a finite interacting system coupled self-consistently to a
noninteracting effective bath. Extensions of DMFT systematically relax
the approximation of locality.\cite{maier_quantum_2005,toschi_dynamical_2007,rubtsov_dual_2008}

In addition to their importance for dynamical mean field theory, quantum
impurity models are extremely important in their own right, since
they appear directly in a variety of problems including nanoscience
(where they are used in describing quantum transport\cite{meir_low-temperature_1993})
and catalysis and surface science (where they can be used to model
the adsorption of molecules on surfaces\cite{brako_slowly_1981}).
The general solution of such impurity models remains a formidable
challenge. Exact solutions\cite{tsvelick_exact_1983} are only available
in specific limits.\cite{andrei_diagonalization_1980,vigman_solution_1982}
A variety of more general semi-analytical\cite{keiter_perturbation_1970,keiter_diagrammatic_1971,pruschke_anderson_1989,georges_hubbard_1992,haule_anderson_2001}
methods have been used successfully. Numerical approaches---all of
which are general in principle, but have different advantages and
limitations in practice---include the numerical renormalization group,\cite{wilson_renormalization_1975,bauer_dynamical_2009}
exact diagonalization,\cite{caffarel_exact_1994} configuration interaction,\cite{zgid_truncated_2012}
hierarchical equations of motion\cite{jin_exact_2008,hou_quantum_2013}
and quantum Monte Carlo.\cite{hirsch_monte_1986,rubtsov_continuous-time_2005,werner_continuous-time_2006,gull_continuous-time_2011}

Continuous time Monte Carlo (CTQMC) algorithms\cite{rubtsov_continuous-time_2005,werner_continuous-time_2006,gull_continuous-time_2011}
are numerically exact and very general in the sense that their computational
complexity is independent of the spectral resolution and band shape,
a property which is particularly crucial in the context of DMFT. Most
algorithms are formulated on the Matsubara axis, and therefore access
to single- and two-particle response functions as measured in experiments
requires an analytic continuation to the real axis. This analytic
continuation problem is ill-posed: small fluctuations in the (input)
Matsubara data, due to Monte Carlo noise, cause large changes in the
output real frequency data and render the direct inversion of the
continuation kernel unreliable for practical purposes. Instead, analytical
continuation algorithms perform the inversion under additional assumptions,
e.g. that the function be described by a small number of zeros and
poles on the complex plane,\cite{haymaker_pade_1970} that it deviate
as little as possible from a `default model' function while being
consistent with the input data within some predetermined error bounds,\cite{jarrell_bayesian_1996}
or that it be as smooth as possible.\cite{prokofev_spectral_2013}

While analytical continuation methods produce spectral functions that
are consistent with Matsubara data, the bias that they introduce on
the real axis is hard to quantify and, as a consequence, controlled
error estimates and confidence intervals on the real axis are not
available even for numerically exact Matsubara data.

Real time methods, on the other hand, embody a controlled way to study
the dynamical properties of quantum impurity models. Monte Carlo impurity
solvers have also been formulated on the Keldysh contour,\cite{muhlbacher_real-time_2008,werner_diagrammatic_2009,schiro_real-time_2009,werner_weak-coupling_2010}
such that analytical continuation is unnecessary (although a combination
of real time and Matsubara techniques may be beneficial\cite{dirks_extracting_2013}).
In the weakly interacting limit, response functions at short times
immediately after a quantum quench have been obtained in this way.\cite{eckstein_thermalization_2009}
Using the newest generation of real time impurity solvers, bold-line\cite{prokofev_bold_2007,prokofev_bold_2008}
CTQMC,\cite{gull_bold-line_2010,gull_numerically_2011} single time
quantities have been obtained at substantially longer times and at
large interaction strength, so that, in combination with reduced dynamics
techniques,\cite{cohen_memory_2011-1} the long time steady state
behavior in the Kondo regime was reached after a quantum quench.\cite{cohen_numerically_2013}

Here we present two ways of obtaining spectral functions from real
time CT-QMC in the strong correlation limit. These methods were recently
used to access nonequilibrium spectral properties;\cite{cohen_greens_2013}
here we present the details of the methodology, focusing for clarity
on the equilibrium aspects and applications. We emphasize, however,
that the methods are equally applicable to nonequilibrium situations.
The first method obtains real frequency spectral functions from two-time
correlation functions which are then Fourier transformed. The second
is based on an auxiliary current formulation, and is more efficient
when one is interested in obtaining spectral steady state or equilibrium
properties. This second method is compatible with the reduced dynamics
technique of Ref.~\onlinecite{cohen_generalized_2013} and with any
numerical solver applicable to transport. It simulates a direct measurement
of the spectral function by coupling auxiliary probes to the system,
and is a generalization of the concept introduced for the wide band
limit by Refs.~\onlinecite{sun_kondo_2001,lebanon_measuring_2001},
which was implemented in this limit within real-time path integral
Monte Carlo.\cite{muhlbacher_anderson_2011} 

In Sec.~\ref{sec:Model} we define the class of models to which our
method is applicable. In Sec.~\ref{sec:Auxiliary-current-method}
we derive a generalized version of the auxiliary current method in
the spirit of previous work, and in Sec.~\ref{sec:Double-probe-scheme}
we introduce a new, fully general scheme compatible with CTQMC. In
Sec.~\ref{sec:Bold-line-Monte-Carlo} we provide an introduction
to the numerically exact real time bold-line CTQMC method; as an example,
we work out the case of the Anderson impurity model, starting from
strong-coupling expansion. In Sec.~\ref{sec:Results} we present
results obtained from an implementation of the scheme for the Anderson
impurity model within bold-line CTQMC, and we explore the limitations
of analytical continuation. Finally, in Sec.~\ref{sec:Summary-and-conclusions},
we summarize our findings and discuss our conclusions.

\section{Model\label{sec:Model}}

We consider a general quantum impurity model described by
\begin{align}
H & =H_{D}+H_{B}+V.\label{eq:hamiltonian}
\end{align}
$H_{D}$, the \emph{dot} Hamiltonian, can be any (generally interacting)
Hamiltonian of the form
\begin{equation}
H_{D}=\sum_{i=1}^{d}\varepsilon_{i}d_{i}^{\dagger}d_{i}+\sum_{ijkl}^{d}U_{ij}^{kl}d_{i}^{\dagger}d_{j}^{\dagger}d_{k}d_{l}+\cdots,\label{eq:dot_hamiltonian_general}
\end{equation}
with $d_{i}^{(\dagger)}$ denoting dot operators. $d$, the total
number of degrees of freedom on the dot, is assumed to be small. $\epsilon_{i}$
describes on-site energy levels, the interaction $U_{ij}^{kl}$ represents
the strength of four-operator (two-body) terms, and the ellipsis all
potential higher order interactions.

The \emph{bath} or \emph{lead} Hamiltonian is: 
\begin{align}
H_{B} & =\sum_{\ell}H_{B\ell},\\
H_{B\ell} & =\sum_{k\in\ell}\varepsilon_{k}a_{k}^{\dagger}a_{k},
\end{align}
 $H_{B}$ is written here as a sum of $N_{B}$ bath terms $H_{B\ell}$
which may be characterized by different dispersions $\varepsilon_{k}$
and different thermodynamic parameters (such as temperature and chemical
potential) entering through the initial conditions. In a molecular
electronics scenario with left and right leads, $N_{B}=2$ and the
index $\ell$ takes the corresponding values $L$ and $R$. $H_{Bl}$
has infinite degrees of freedom described by lead operators $a_{k}^{(\dagger)}$
and a lead dispersion $\varepsilon_{k}$ but is noninteracting.

The third term is the \emph{hybridization} Hamiltonian $V$, which
describes population transfer between the dot and leads. Here we assume
it to have the bilinear form 
\begin{align}
V & =\sum_{\ell}V_{\ell},\\
V_{\ell} & =\sum_{i=1}^{d}\sum_{k\in\ell}\left(t_{ik}a_{k}^{\dagger}d_{i}+t_{ik}^{*}d_{i}^{\dagger}a_{k}\right)
\end{align}
characterized by a hopping $t_{ik}$ from dot to lead. The coupling
densities $\Gamma_{\ell ij}(\omega),$ defined as $\Gamma_{\ell ij}\left(\omega\right)=2\pi\sum_{k\in\ell}t_{ik}^{*}t_{jk}\delta\left(\omega-\varepsilon_{k}\right)$,
fully define the properties of bath and hybridizations for the purpose
of this paper. We will assume $\Gamma_{\ell ij}=\delta_{ij}\Gamma_{\ell}$
to be diagonal in dot orbital space but allow it to differ for each
lead, such that $\Gamma_{\ell}\left(\omega\right)\equiv2\pi\sum_{k\in\ell}\left|t_{k}\right|^{2}\delta\left(\omega-\varepsilon_{k}\right)$.
General coupling density matrices will be discussed briefly in Sec.~\ref{sec:Double-probe-scheme}. 

Information about thermodynamic parameters, e.g. the lead chemical
potential or the lead temperature, is encapsulated in the initial
conditions of the system. For example, fermionic leads at an inverse
temperature $\beta_{\ell}$ and a chemical potential $\mu_{\ell}$
are described by an initial Fermi Dirac distribution $f_{\ell}\left(\omega\right)\equiv\frac{1}{1+e^{\beta_{\ell}\left(\omega-\mu_{\ell}\right)}}$
.

\section{Auxiliary current method\label{sec:Auxiliary-current-method}}

\begin{figure}
\includegraphics[width=8.6cm]{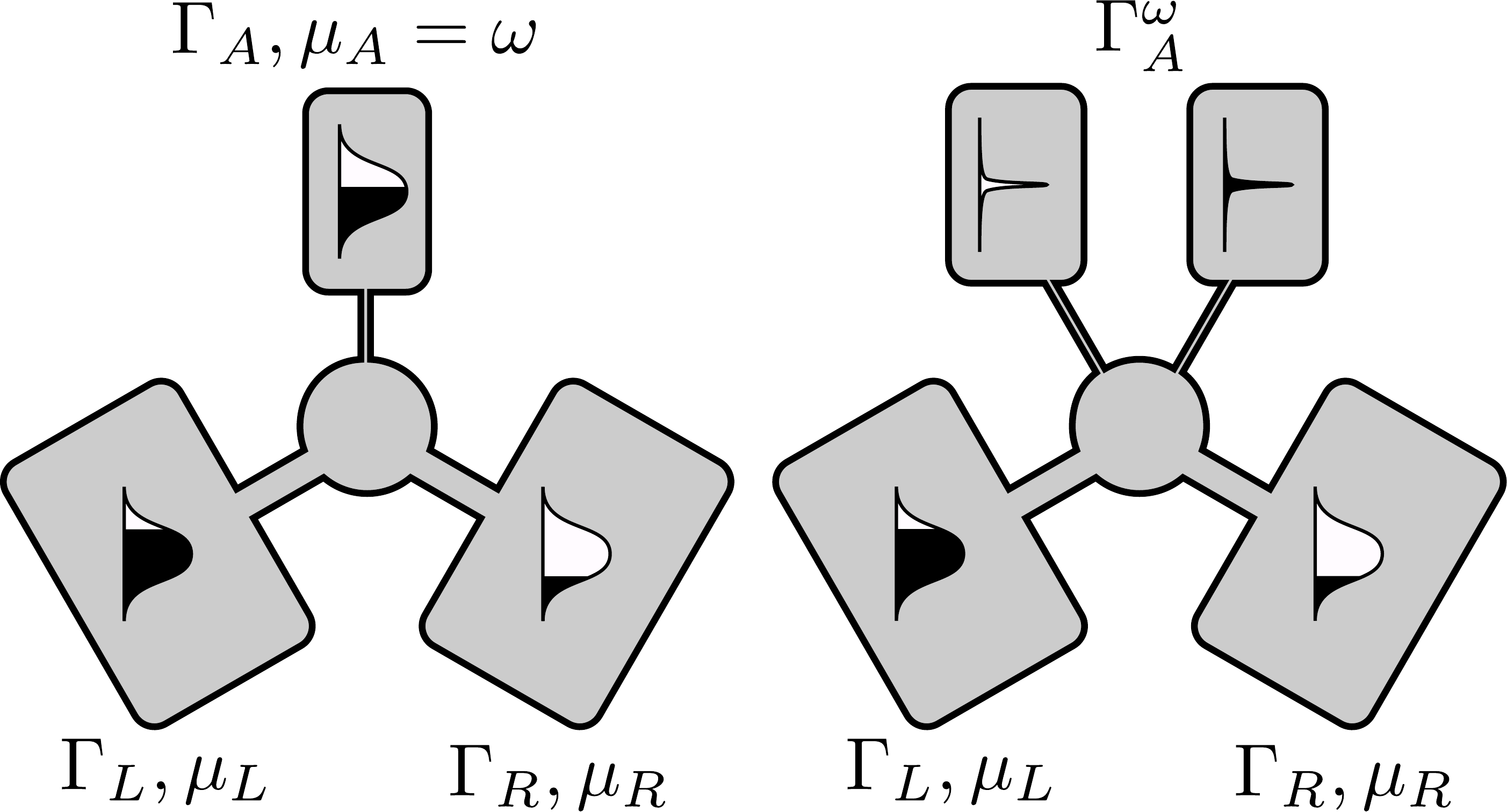}\caption{Left: an illustration of the single probe, wide-band auxiliary current
setup. Right: the double probe, narrow-band variation of the auxiliary
current formalism. The dot is depicted as the central circle, and
is coupled by thick (thin) lines to the physical (auxiliary) reservoirs.
The curved region within each reservoir sketches the shape and filling
of its coupling density. Whereas in the single probe apparatus a single
high-bandwidth auxiliary reservoir is coupled to the dot and the integral
of the spectral function is obtained by measuring the auxiliary current
while varying its chemical potential, in the double probe scheme (suggested
as a theoretical tool rather than an experiment) the spectral function
is obtained without the need for a derivative by attaching two low-bandwidth
leads, of which one is empty and the other is full.\label{fig:apparatus}}
\end{figure}

In their seminal 1992 paper, Meir and Wingreen showed that if initial
correlations can be neglected, the steady-state current $I_{\ell}$
out of lead $\ell$ can be written as:\cite{meir_landauer_1992}
\begin{align}
I_{\ell} & =\frac{ie}{2h}\int\mathrm{d}\omega\label{eq:current_per_lead}\\
 & \,\times\mathrm{Tr}\left\{ \Gamma_{\ell}\left(\omega\right)\left[f_{\ell}\left(\omega\right)\left[G^{r}\left(\omega\right)-G^{a}\left(\omega\right)\right]+G^{<}\left(\omega\right)\right]\right\} .\nonumber 
\end{align}
Here, $e$ and $h$ are the electron charge and Planck's constant
(both are set to one from here onward); $f_{\ell}$ is the initial
occupation of lead $\ell$; and $G^{r}$,$G^{a}$ and $G^{<}$ are
the dot's retarded, advanced and lesser Green's functions, respectively.
The difference between the retarded and advanced Green's function
is proportional to the dot's spectral function: 
\begin{equation}
\left[G^{r}\left(\omega\right)-G^{a}\left(\omega\right)\right]=2i\Im\left\{ G^{r}\left(\omega\right)\right\} =-2\pi iA\left(\omega\right).\label{eq:A_from_Gs}
\end{equation}
As previous authors have pointed out,\cite{sun_kondo_2001,lebanon_measuring_2001}
if the $\Gamma_{\ell}(\omega)=\Gamma_{l}$ are independent of energy
and the orbital structure of $A\left(\omega\right)$ is diagonal at
all frequencies such that it can be treated as a scalar (i.e. $\langle d_{i}^{\dagger}d_{j}\rangle=0$
for $i\neq j$), then $A\left(\omega\right)$ can be obtained by coupling
the system to an additional auxiliary lead with index $\ell=A$. Here
we generalize this approach to the case where the $\Gamma_{\ell}(\omega)$
are not energy independent but proportional: $\Gamma_{\ell}\left(\omega\right)=\lambda_{\ell}\Gamma\left(\omega\right)$
and $\Gamma_{A}\left(\omega\right)=\eta\Gamma\left(\omega\right)$
(note that the next section introduces a generalized scheme where
these assumptions are not required). The extended system, illustrated
on the left panel of Fig.~\ref{fig:apparatus}, is described by the
following modifications to the Hamiltonian: 
\begin{align}
H_{B} & \rightarrow H_{B}+H_{BA},\\
V & \rightarrow V+V_{A},\\
H_{BA} & \equiv\sum_{k\in A}\varepsilon_{k}a_{k}^{\dagger}a_{k},\\
V_{A} & \equiv\sum_{i}\sum_{k\in A}\left(t_{ik}a_{k}^{\dagger}d_{i}+t_{ik}^{*}d_{i}^{\dagger}a_{k}\right),
\end{align}
with $\epsilon_{k}$and $t_{ik}$ chosen such that $\Gamma_{A}(\omega)=\eta\Gamma(\omega)$.
The index $A$ denotes the auxiliary lead. Following Refs.~\onlinecite{sun_kondo_2001,lebanon_measuring_2001},
we use the conservation of current $\sum_{\ell}I_{\ell}+I_{A}=0$
along with Eq.~\prettyref{eq:current_per_lead} and \prettyref{eq:A_from_Gs}
to construct an expression for $I_{A}$ in which $G^{<}$ no longer
appears:
\begin{align}
I_{A} & =I_{A}-\frac{\eta}{\sum_{\ell}\lambda_{\ell}+\eta}\left(\sum_{\ell}I_{\ell}+I_{A}\right)\\
 & =\pi\int\mathrm{d}\omega\mathrm{Tr}\left\{ \Gamma\left(\omega\right)A\left(\omega\right)\right\} \\
 & \,\times\left\{ \frac{\eta}{\sum_{\ell}\lambda_{\ell}+\eta}\left[f_{A}\left(\sum_{\ell}\lambda_{\ell}\right)-\sum_{\ell}\lambda_{\ell}f_{\ell}\right]\right\} .\nonumber 
\end{align}
In the limit where $\eta\rightarrow0$, system properties such as
the Green's functions are unaffected by the presence and properties
of the auxiliary lead. Therefore, the derivative of the current $I_{A}$
with respect to the chemical potential in the auxiliary lead, $\frac{\mathrm{d}I_{A}}{\mathrm{d}\mu_{A}}$,
only contains contributions from $f_{A}$:
\begin{align}
\lim_{\eta\rightarrow0}\frac{\mathrm{d}}{\mathrm{d}\mu_{A}}I_{A} & =\lim_{\eta\rightarrow0}\pi\int\mathrm{d}\omega\mathrm{Tr}\left\{ \Gamma\left(\omega\right)A\left(\omega\right)\right\} \\
 & \,\times\left\{ \frac{\eta}{\sum_{\ell}\lambda_{\ell}+\eta}\left[\frac{\mathrm{d}f_{A}}{\mathrm{d}\mu_{A}}\left(\sum_{\ell}\lambda_{\ell}\right)\right]\right\} .\nonumber 
\end{align}
With our assumption that $\Gamma$ and $A$ are diagonal, the trace
of $\Gamma A$ over the impurity degrees of freedom is a product over
their elements (up to a factor of $d$, which we will ignore)\emph{.}
If lead $A$ is maintained at low enough temperature, we can also
set $\lim_{\beta_{A}\rightarrow\infty}\frac{\mathrm{d}f_{A}\left(\omega\right)}{\mathrm{d}\mu_{A}}=-\delta\left(\omega-\mu_{A}\right)$,
which allows us to perform the integration and obtain
\begin{align}
A\left(\mu_{A}\right) & \underset{{\scriptstyle \eta\rightarrow0}}{=}-\frac{1}{\pi}\frac{1}{\eta}\Gamma^{-1}\left(\mu_{A}\right)\frac{\sum_{\ell}\lambda_{\ell}}{\sum_{\ell}\lambda_{\ell}+\eta}\frac{\mathrm{d}I_{A}}{\mathrm{d}\mu_{A}}\label{eq:A_from_aux_current_old}\\
 & =-\frac{\sum_{\ell}\Gamma_{\ell}\left(\mu_{A}\right)}{\sum_{\ell}\Gamma_{\ell}\left(\mu_{A}\right)+\Gamma_{A}\left(\mu_{A}\right)}\frac{1}{\pi}\Gamma_{A}^{-1}\left(\mu_{A}\right)\frac{\mathrm{d}I_{A}}{\mathrm{d}\mu_{A}}.\label{eq:A_from_aux_current_old_no_eta}
\end{align}
This result is identical to the one of Refs.~\onlinecite{sun_kondo_2001,lebanon_measuring_2001},
except in that $\Gamma$ is allowed to have an arbitrary energy dependence.
The approach outlined here is, in general, limited to systems with
only diagonal dot correlation and hybridization functions.

Eq.~\prettyref{eq:A_from_aux_current_old} and \prettyref{eq:A_from_aux_current_old_no_eta}
allow sampling the spectral function by measuring the current flow
between the dot and a weakly coupled auxiliary lead at a variety of
chemical potentials. This could also in principle be done in an experimental
setting by varying the gate voltage of an auxiliary lead, and this
is perhaps the greatest advantage of this variant of the method. However,
the setup is only useful in cases where $\Gamma\left(\omega\right)$
is not suppressed. Near regions where $\Gamma\left(\omega\right)$
is small, $\frac{\mathrm{d}I_{A}}{\mathrm{d}\mu_{A}}$ must be very
small in order for $A$ not to diverge, and the numerical computation
or measurement of the derivative must be performed to a very high
accuracy. For this reason, the presence of noise makes the numerical
computation of this derivative impractical within a Monte Carlo simulation,
at least for cases where one is interested in behavior in or near
a gap in the lead's spectrum.

\section{Double probe scheme\label{sec:Double-probe-scheme}}

In order to bypass the limitation outlined above and the need to compute
derivatives, we propose an alternative scheme. The current in the
auxiliary lead is (see\prettyref{eq:current_per_lead})
\begin{align}
I_{A} & =\frac{ie}{2h}\int\mathrm{d}\omega\label{eq:current_aux_lead}\\
 & \,\times\mathrm{Tr}\left\{ \Gamma_{A}\left(\omega\right)\left[f_{A}\left(\omega\right)\left[G^{r}\left(\omega\right)-G^{a}\left(\omega\right)\right]+G^{<}\left(\omega\right)\right]\right\} .\nonumber 
\end{align}
In the limit $\Gamma_{A}\rightarrow0$ the partial derivative of this
equation with respect to $\mu_{A}$ immediately yields Eq.~\prettyref{eq:A_from_aux_current_old_no_eta}
without the factor $\frac{\sum_{\ell}\Gamma_{\ell}}{\sum_{\ell}\Gamma_{\ell}+\Gamma_{A}}\simeq1-\frac{\Gamma_{A}}{\sum_{\ell}\Gamma_{\ell}}=1-\frac{\eta}{\sum_{\ell}\lambda_{\ell}}$,
the higher order terms of which constitute corrections in powers of
$\frac{\eta}{\sum_{\ell}\lambda_{\ell}}$. Our form ignores the secondary
effects of the auxiliary lead on the currents through other leads,
and only becomes accurate in the limit of an auxiliary lead that is
weakly coupled to the system ($\eta$ small), a criterion that can
be verified \emph{a posteriori} by varying $\eta$. Additionally,
no assumptions regarding the relationship or proportionality between
$\Gamma_{A}$ and $\left\{ \Gamma_{\ell}\right\} $ need to be made.
This freedom allows the experimentally unrealistic but numerically
convenient choice of introducing \emph{two} auxiliary leads, one empty
and one full, set up in such a way that they are coupled to the system
at only a single frequency $\omega^{\prime}$:

\begin{eqnarray}
f_{A}^{i}\left(\omega\right) & = & \begin{cases}
0: & i=0,\\
1: & i=1,
\end{cases}\\
\Gamma_{A}^{\omega^{\prime}}\left(\omega\right) & = & \eta\delta\left(\omega-\omega^{\prime}\right).\label{Gamma_a_delta}
\end{eqnarray}
Using these definitions, we define two auxiliary currents
\begin{eqnarray}
I_{A}^{0}\left(\omega^{\prime}\right) & = & \frac{i}{2}\eta\left\{ G^{<}\left(\omega^{\prime}\right)\right\} ,\label{eq:I0}\\
I_{A}^{1}\left(\omega^{\prime}\right) & = & \frac{i}{2}\eta\left\{ \left[G^{r}\left(\omega^{\prime}\right)-G^{a}\left(\omega^{\prime}\right)\right]+G^{<}\left(\omega\right)\right\} \label{eq:I1}\\
 & = & \eta\frac{\pi}{2}A\left(\omega^{\prime}\right)+I_{A}^{0}\left(\omega^{\prime}\right).
\end{eqnarray}
The assumption made here, as in Ref.~\onlinecite{sun_kondo_2001,lebanon_measuring_2001}
and the preceding section, is that $\eta$ is small enough that the
properties of the auxiliary lead have a negligible effect on the physical
properties of the full system. We note in passing that this same idea
also allows for the use of a single wide-bandwidth auxiliary lead
in the case where the coupling densities are \emph{not} proportional.

Two independent calculations, one for an empty lead and a second one
for a full lead, provide a setup (see right side of Fig.~\ref{fig:apparatus})
with which the dot spectral function can be obtained. Restoring physical
constants for a moment,
\begin{equation}
A\left(\omega\right)=\lim_{\eta\rightarrow0}-\frac{2h}{e\pi\eta}\left[I_{A}^{1}\left(\omega\right)-I_{A}^{0}\left(\omega\right)\right].\label{eq:A_double_probe_scheme}
\end{equation}
This analytically exact result is not restricted to any particular
way of solving the impurity model, and is therefore usable within
any formalism where one has access to currents. Apart from quantum
Monte Carlo this includes hierarchical equation of motion methods.
Intuitively, the introduction of two leads of opposite populations
simultaneously should correct current conservation to some degree
(since the coupling densities are no longer proportional, an exact
statement is hard to make), but they can also be connected one at
a time at the price of a small loss in accuracy, as long as $\eta$
is taken to be small enough.

Equations Eq.~\prettyref{eq:I0} and Eq.~\prettyref{eq:I1} also
provide direct access to the lesser Green's function $G^{<}\left(\omega\right)$,
providing the full information about single particle correlations
even in nonequilibrium situations. Finally, since $\Gamma_{A}^{\omega^{\prime}}\left(\omega\right)$
can be chosen to have any dot orbital matrix structure, it is straightforward
to find a choice that allows extracting individual elements of $A_{ij}\left(\omega\right)$.

Since no numerical derivative needs to be taken, the double probe
scheme performs better within a Monte Carlo simulation, in addition
to providing more information (such as $G^{<}$). Any practical implementation
will need to approximate the delta function in Eq.~\ref{Gamma_a_delta}
by a smooth numerical approximation. In this work, we approximate
$\Gamma_{A}^{\omega'}$ with Gaussians with an amplitude and width
small enough that the results are independent of them within the numerical
precision bounds and frequency resolution we need (a good rule of
thumb is to start the amplitude and width at least an order of magnitude
or two below any other energy scale in the problem; one then verifies
convergence in both parameters of the Gaussian by systematically varying
them). The picture which emerges is physically intuitive: we probe
the system using a pair of virtual leads, one of which is empty of
electrons while the one other is full. The leads are coupled very
weakly and only within a narrow frequency range. By measuring the
rate at which electrons are injected into one lead and extracted from
the other, one obtains information about the average density of single-electron
excitations within the frequency range probed.

The auxiliary current method by construction contains two-time information,
but strictly speaking this only corresponds to the two-time correlation
function in a certain adiabatic limit (where the relaxation of the
auxiliary current to its final value is much faster than that of the
system's dynamical timescales). However, the method has a straightforward
extension to the handling of two-time correlation functions; this
can be seen by considering eq. (15) in Ref.~\onlinecite{jauho_time-dependent_1994},
which connects the current through any lead $\ell$ to the two-time
Green's function $G^{r}\left(t,t_{1}\right)$ by way of a generalized
coupling density $\Gamma_{\ell}\left(\omega,t_{1},t\right)$. By considering
this expression, it is easy to see how a generalized, time-dependent
auxiliary coupling density might be devised, which fully and rigorously
probes the full two-time correlation functions. This is beyond the
scope of the present work.

\section{Bold-line Monte Carlo\label{sec:Bold-line-Monte-Carlo}}

The auxiliary current formalism is only useful in combination with
a method for computing currents through interacting quantum dots.
This requires the solution of a correlated quantum many-body problem,
and therefore a numerical method. Several approaches exist. In this
section we present a detailed description of the real-time bold-line
continuous-time hybridization expansion quantum Monte Carlo algorithm
introduced in Ref.~\onlinecite{gull_numerically_2011}. We focus
in particular on the observables required for obtaining two-time correlation
functions but, for the sake of simplicity, limit ourselves to the
single-orbital Anderson impurity model.

The method is based on a stochastic summation of all diagrams containing
partially summed (`bold' rather than `bare') propagator lines and
vertex functions. So far, these partially summed propagators have
come from the non-crossing or one-crossing approximations (NCA or
OCA\cite{cohen_numerically_2013}) but the method is more general:
In a first, quasi-analytic step, an underlying diagrammatic approximation
selecting some (but not all) diagrams is chosen, and propagators within
that approximation are obtained analytically or via the solution of
a set of coupled integral equations. In a second step, all corrections
to the propagators are summed up using a stochastic Monte Carlo procedure,
so that the resulting sum contains all diagrams and therefore becomes
numerically exact. The precise choice of the underlying approximation
determines the speed of convergence to the exact result, the statistical
uncertainty, and the feasibility of the method for any given system,
but has no effect on the final answer if convergence is attained.
The imaginary time version of bold-line CTQMC has been introduced
in Ref.~\onlinecite{gull_bold-line_2010}. The formulation on the
Keldysh contour and the advantages of bold-line CTQMC over bare CTQMC
in the non-equilibrium context have been discussed and benchmarked
in Ref.~\onlinecite{gull_numerically_2011}, and an extension to
OCA and to a reduced dynamics formulation has appeared in Ref.~\onlinecite{cohen_numerically_2013}.

We consider the single-orbital Anderson impurity model, where we set
$d=2$ in Eq.~\prettyref{eq:dot_hamiltonian_general} such that there
may be up to two electrons having opposite spin indices $\uparrow$
and $\downarrow$ on the dot simultaneously. The dot Hamiltonian becomes
\begin{equation}
H_{D}=\sum_{\sigma=\uparrow,\downarrow}\varepsilon_{\sigma}d_{\sigma}^{\dagger}d_{\sigma}+Ud_{\uparrow}^{\dagger}d_{\uparrow}d_{\downarrow}^{\dagger}d_{\downarrow},
\end{equation}
the lead Hamiltonian

\begin{align}
H_{B} & =\sum_{\sigma=\uparrow,\downarrow}\sum_{k\in\ell_{\sigma}}\varepsilon_{k}a_{k}^{\dagger}a_{k},
\end{align}
and the dot-lead hybridization 
\begin{equation}
V_{\ell}=\sum_{\sigma\in\uparrow,\downarrow}\sum_{k\in\ell_{\sigma}}\left(t_{k}a_{k}^{\dagger}d_{\sigma}+t_{k}^{*}d_{\sigma}^{\dagger}a_{k}\right).
\end{equation}

We first derive a version of the bare real time hybridization expansion
by introducing a perturbation theory written in terms of many-body
atomic state propagators. The expectation value of an operator $A$
at time $t$ in the interaction picture with respect to $H_{0}\equiv H-V$
is
\begin{align}
\left\langle A\left(t\right)\right\rangle  & =\mathrm{Tr}\left\{ \rho U^{\dagger}\left(t\right)A_{I}\left(t\right)U\left(t\right)\right\} .
\end{align}
Here $\rho=\rho_{D}\otimes\rho_{B}$ is the initial density matrix.
We assume that it can be factorized into dot and lead parts (this
assumption can be relaxed by adding an imaginary branch to the contour).
$U\left(t\right)=e^{iH_{0}t}e^{-iHt}$ is the interaction picture
propagator, and interaction picture operators are denoted by a subscript
$I$ and given by $A_{I}\left(t\right)=e^{iH_{0}t}Ae^{-iH_{0}t}$.
We then expand $U$ and $U^{\dagger}$ in the form\cite{mahan_many-particle_1990}
\begin{align}
U\left(t\right) & =\sum_{n=0}^{\infty}\left(-i\right)^{n}\int_{0}^{t}\mathrm{d}t_{1}\int_{0}^{t_{1}}\mathrm{d}t_{2}...\int_{0}^{t_{n-1}}\mathrm{d}t_{n}\nonumber \\
 & \,\times V_{I}\left(t_{1}\right)...V_{I}\left(t_{n}\right),\label{eq:interaction_propagator_series}
\end{align}
thus obtaining an infinite series of terms. It is convenient to imagine
that interactions coming from $U\left(t\right)$ exist on a \emph{forward}
branch while those coming $U^{\dagger}\left(t\right)$ are on a \emph{backward}
branch; the union of these two contours forms the real part of the
Keldysh contour. Notably, since in our choice of model and expansion
$H_{0}$ is interacting, Wick's theorem does not hold and one cannot
at this point apply standard diagrammatic tools.

The interaction form of the Hybridization term $V_{I}(t)$ is
\begin{align}
V_{I\ell}\left(t\right) & =\sum_{\sigma\in\uparrow,\downarrow}\sum_{k\in\ell}\left\{ t_{\sigma k}e^{i\left(\varepsilon_{\sigma}+Ud_{\bar{\sigma}}^{\dagger}d_{\bar{\sigma}}-\varepsilon_{\sigma k}\right)t}a_{\sigma k}^{\dagger}d_{\sigma}\right.\nonumber \\
 & \,\left.+t_{\sigma k}^{*}e^{-i\left(\varepsilon_{\sigma}+Ud_{\bar{\sigma}}^{\dagger}d_{\bar{\sigma}}-\varepsilon_{\sigma k}\right)t}d_{\sigma}^{\dagger}a_{\sigma k}\right\} ,\label{eq:V_interaction_picture}
\end{align}
with $\bar{\sigma}$ denoting the spin opposite of $\sigma$. All
terms of $V_{\ell}$ change the state of the dot by adding or removing
one electron from it, while time evolution with $H_{0}$ leaves the
dot state invariant. It is therefore convenient to group certain sets
of contributions together in \emph{dressed ($G$) }and \emph{bare
($G^{\left(0\right)})$ }propagators 

\begin{align}
G_{\alpha\beta}\left(t\right) & \equiv\left\langle \alpha\right|\mathrm{Tr}_{B}\left\{ \rho e^{-iHt}\right\} \left|\beta\right\rangle ,\label{eq:propagator_definition}\\
G_{\alpha\alpha}^{(0)}\left(t\right) & \equiv\left\langle \alpha\right|\mathrm{Tr}_{B}\left\{ \rho e^{-iH_{0}t}\right\} \left|\alpha\right\rangle ,\label{eq:bare_propagator_definition}
\end{align}
between the many-body states $\alpha$ and $\beta$ on the dot, with
the leads traced out. Both atomic state propagators and correlation
functions are often referred to as Green's functions (in addition,
the retarded and advanced Green's functions are often referred to
as propagators), but the objects defined here differ from correlation
functions in several important ways---most notably, they have an exponentially
larger dimensionality, since they contain matrix elements between
pairs of many-body states $\alpha,\beta$ rather than single-particle
level indices. Diagrammatically, a propagator over a segment of one
branch of the Keldysh contour represents all diagrams having all their
interactions contained in said segment, that begin and end with the
index states.

\begin{figure}
\includegraphics[width=8.6cm]{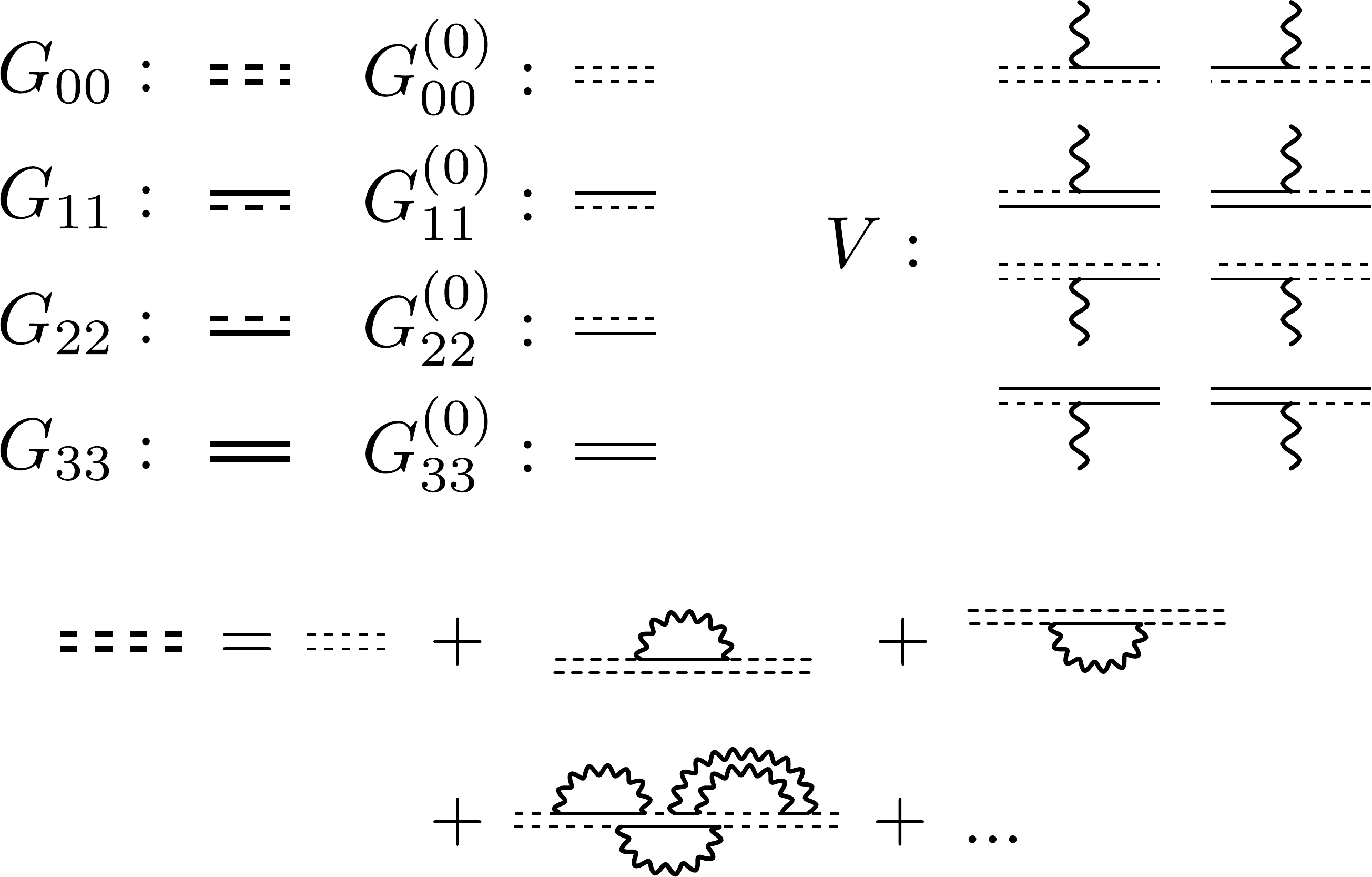}\caption{The elements of atomic state propagator diagrams. Upper left: full
and bare propagator lines. Upper right: interaction (`hybridization')
vertices. Bottom: examples of low order diagrams. The upper line represents
spin up, the lower line spin down. Solid lines denote occupied orbitals,
dashed lines empty orbitals. Thick lines denote dressed propagators,
thin lines bare propagators. Wiggly lines denote the ejection of an
electron to the bath or its propagation back to the dot.\label{fig:propagator_elements}}
\end{figure}

For the Anderson model, the bare atomic state propagators (Eq.~\ref{eq:bare_propagator_definition})
are
\begin{equation}
G_{\alpha\beta}^{\left(0\right)}\left(t\right)=\Phi\left(t\right)\delta_{\alpha\beta}e^{-i\varepsilon_{\alpha}t}.\label{eq:G_0_value}
\end{equation}
The factor $\Phi\left(t\right)\equiv\mathrm{Tr}_{B}\left\{ \rho_{B}e^{-iH_{B}t}\right\} $
is independent of the dot state and factors out of all expressions;
for physical quantities it is always exactly canceled by complementary
contributions from the two branches of the Keldysh contour: It can
therefore safely be ignored and will be neglected from here onwards.

Since the trace over the bath is zero unless the same number of creation
and annihilation operators for each spin occurs, the full atomic state
propagator is also diagonal for the Anderson impurity model. It can
be expanded in terms of bare propagators using Eq.~\ref{eq:interaction_propagator_series}.
The first non-vanishing contribution is:
\begin{align}
G_{\alpha\alpha}\left(t\right) & =\left\langle \alpha\right|\mathrm{Tr}_{B}\left\{ \rho e^{-iH_{0}t}U\left(t\right)\right\} \left|\alpha\right\rangle \\
 & =\left\langle \alpha\right|\mathrm{Tr}_{B}\left\{ \rho e^{-iH_{0}t}\right\} \left|\alpha\right\rangle \label{eq:secondorder}\\
 & \,+\left(-i\right)^{2}\int_{0}^{t}\mathrm{d}t_{1}\int_{0}^{t_{1}}\mathrm{d}t_{2}\nonumber \\
 & \,\times\left\langle \alpha\right|\mathrm{Tr}_{B}\left\{ \rho e^{-iH_{0}t}V_{I}\left(t_{1}\right)V_{I}\left(t_{2}\right)\right\} \left|\alpha\right\rangle \nonumber \\
 & \,+...\nonumber 
\end{align}
Using Eq.~\ref{eq:V_interaction_picture} and Eq.~\ref{eq:G_0_value},
this can be written as
\begin{align}
G_{\alpha\alpha}\left(t-t^{\prime}\right) & =G_{\alpha\alpha}^{\left(0\right)}\left(t-t^{\prime}\right)\\
 & \,+\sum_{\beta}\int_{0}^{t}\mathrm{d}t_{1}\int_{0}^{t_{1}}\mathrm{d}t_{2}\nonumber \\
 & \,\times G_{\alpha\alpha}^{\left(0\right)}\left(t-t_{1}\right)G_{\beta\beta}^{\left(0\right)}\left(t_{1}-t_{2}\right)G_{\alpha\alpha}^{\left(0\right)}\left(t_{2}-t^{\prime}\right)\nonumber \\
 & \,\times\Delta_{\alpha\alpha}^{\beta}\left(t_{1}-t_{2}\right)\nonumber \\
 & \,+...,\nonumber 
\end{align}
where 
\begin{align}
\Delta_{\alpha\alpha}^{\beta}\left(t_{1}-t_{2}\right) & \equiv\langle\alpha|d^{\dagger}|\beta\rangle\langle\beta|d|\alpha\rangle\label{eq:hybridization_function_definition}\\
 & \times\sum_{k\in\ell}\left|t_{k}\right|^{2}\mathrm{Tr}_{B}\left\{ a_{Ik}\left(t_{1}\right)a_{Ik}^{\dagger}\left(t_{2}\right)\right\} \nonumber \\
 & +\langle\alpha|d|\beta\rangle\langle\beta|d^{\dagger}|\alpha\rangle\nonumber \\
 & \times\sum_{k\in\ell}\left|t_{k}\right|^{2}\mathrm{Tr}_{B}\left\{ a_{Ik}^{\dagger}\left(t_{1}\right)a_{Ik}\left(t_{2}\right)\right\} \nonumber 
\end{align}
defines the \emph{hybridization function} (note that spin indices
have been suppressed for brevity). The hybridization function $\Delta_{l}(t_{1,},t_{2})=\sum_{k\in\ell}\left|t_{k}\right|^{2}\mathrm{Tr}_{B}\left\{ a_{Ik}^{\dagger}\left(t_{1}\right)a_{Ik}\left(t_{2}\right)\right\} $
for each lead $\ell$ can be expressed in terms of the coupling densities
$\Gamma_{\ell}$ of that lead and its initial occupation probability,
both of which are obtained from the lattice Green's function within
DMFT. It is $\Delta_{\ell}^{>}$ if $t_{1}$ appears before $t_{2}$
on the Keldysh contour, and $\Delta_{\ell}^{<}$ otherwise, with

\begin{align}
\Delta_{\ell}^{<}(t_{1},t_{2}) & =-2i\int_{-\infty}^{\infty}\frac{d\omega}{2\pi}e^{-i\omega(t_{1}-t_{2})}\Gamma_{\ell}(\omega)f(\omega-\mu_{\ell}),\\
\Delta_{\ell}^{>}(t_{1},t_{2}) & =2i\int_{-\infty}^{\infty}\frac{d\omega}{2\pi}e^{-i\omega(t_{1}-t_{2})}\Gamma_{\ell}(\omega)[1-f(\omega-\mu_{\ell})].
\end{align}

This second order term contains all the elements occurring in the
expansion, namely the bare propagators and hybridization lines. A
graphical representation is shown in Fig.~\ref{fig:propagator_elements}:
in the upper left panel, atomic state propagators for a given state
are represented by a pair of lines; the upper one representing spin
up, the lower one spin down. Dotted lines represent empty states and
solid lines represent occupied states, such that the full set of $2^{N}$
states is represented by $N$ lines. Additionally, thick (or bold)
lines represent full propagators while thin lines stand for bare propagators.
Full propagators are built out of all possible combinations of bare
propagators and interaction vertices (shown to the upper right), with
all interaction times integrated over. Every hybridization line connects
two interactions, and each such line comes with the minus sign of
Eq.~\ref{eq:secondorder}. The second order contribution explicitly
written above includes both the $2{}^{\mathrm{nd}}$ and $3^{\mathrm{rd}}$
term in the expansion of $G_{00}$ shown in the bottom part of Fig.~\ref{fig:propagator_elements},
where an additional $8^{\mathrm{th}}$ order diagram is also shown.

The essence of the CTQMC method\cite{muhlbacher_real-time_2008,werner_diagrammatic_2009,gull_continuous-time_2011,rubtsov_continuous-time_2005}
is that one can evaluate the sum of all diagrams by sampling them
stochastically using the Metropolis algorithm. In particular, the
`continuous time' descriptor implies that, unlike in previous algorithms,\cite{hirsch_monte_1986}systematic
Trotter time-discretization errors are absent.\cite{gull_performance_2007}
In the hybridization expansion continuous-time algorithm,\cite{werner_continuous-time_2006,gull_continuous-time_2011}
the stochastic sampling of diagrams is done by insertion of pairs
of dot operators and hybridization lines: Namely, in each Monte Carlo
step, a hybridization line and two dot operators are created, destroyed,
or moved. This set of updates can be shown to respect ergodicity,
and moves can be weighed in such a way that detailed balance is also
maintained.\cite{gull_continuous-time_2011} Since the leads are noninteracting
both initially and under propagation by $H_{0}$, Wick's theorem is
respected for lead (though not for the dot) operators, and it is possible
to efficiently sum large sets of diagrams having the same \emph{path}---dot
state as a function of time---in the form of determinants, significantly
reducing the sign problem.\cite{werner_continuous-time_2006}

\begin{figure}
\includegraphics[width=8.6cm]{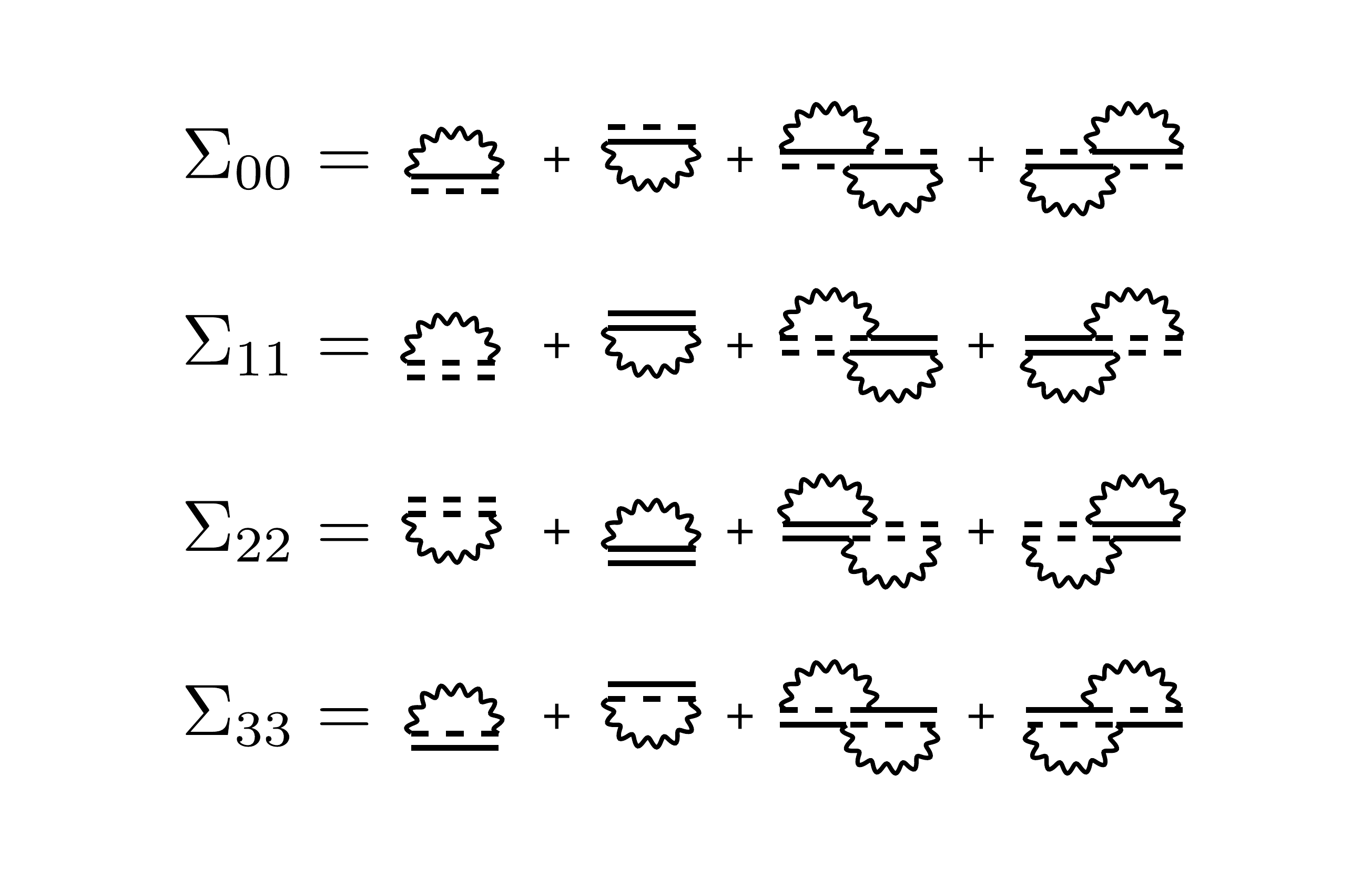}\caption{The matrix elements of the NCA self energy (two terms on the left
with a single hybridization line) and of the OCA self energy (all
four terms) in diagrammatic form.\label{fig:dyson_equation}}
\end{figure}

In bold-line CTQMC, rather than summing each diagram explicitly, one
first obtains an approximation for the atomic state propagator that
is better than the bare atomic state propagator by dressing it using
some ansatz for the self energy. If one chooses a self-consistent
approximation such as the NCA or OCA, this allows for the relatively
inexpensive computation of a renormalized, or bold line, propagator
which contains an infinite (though still partial) set of diagrams.
This is not the full propagator, but often represents a reasonable
approximation to it. It is then possible to write all \emph{additional}
diagrams in terms of this new bold propagator. To avoid double counting
diagrams, one must take into account only diagrams which do not contain
parts already summed within the underlying approximation. In a Monte
Carlo algorithm this is feasible if one rejects any update that leads
to an unwanted diagram, while ascertaining that still all diagrams
are generated (thus maintaining ergodicity). Each diagram in the bare
expansion is contained in exactly one bold diagram, which has the
same number of or fewer vertices. Since the original expansion was
convergent, the sampling process of all diagrams gives the same exact
answer---however, if the renormalized propagators contain a large
proportion of the most important contributions, convergence occurs
at lower diagram order, is substantially faster, and the sign problem
is greatly alleviated.\cite{gull_numerically_2011}

The NCA approximation to the propagator is described pictorially in
the first two terms of Fig.~\ref{fig:dyson_equation} with one hybridization
line (see also Ref.~\onlinecite{cohen_numerically_2013}, which uses
a more compact notation). The OCA also includes the two additional
diagrams with two hybridization lines. The auxiliary state self energy
$\Sigma_{\alpha\alpha}$ is inserted into the causal Dyson equation
\begin{align}
G_{\alpha\alpha}\left(t-t^{\prime}\right) & =G_{\alpha\alpha}^{(0)}\left(t-t^{\prime}\right)\\
 & \,+\int_{t^{\prime}}^{t}\mathrm{d}t_{1}\int_{t^{\prime}}^{t_{1}}\mathrm{d}t_{2}\times\nonumber \\
 & \, G_{\alpha\alpha}^{(0)}\left(t-t_{1}\right)\Sigma_{\alpha\alpha}\left(t_{1}-t_{2}\right)G_{\alpha\alpha}\left(t_{2}-t^{\prime}\right)\nonumber 
\end{align}
and solved self-consistently. For Hamiltonians without an explicit
time dependence, like the Anderson impurity Hamiltonian considered
here, the propagator is a function of only time differences even outside
of steady state conditions. This can be seen directly from its definition
in Eq.~\ref{eq:propagator_definition}. In terms of numerical effort
the cost of solving the Dyson equation is therefore negligible even
for very long propagation times, within both the NCA and the OCA.

Bold diagrams no longer have the determinant structure of bare diagrams,
since some permutations of lead operators are contained in the underlying
partial summation, while most are not. It is therefore necessary to
explicitly treat the sum of individual hybridization lines. In practice,
cancellation effects of diagrams with the same dot operator configuration
but different hybridization lines are large, so that taking into account
all legal hybridization line contractions of a given dot operator
configuration at once, similar to computing the determinant, is preferable
to sampling them individually. The cost of explicitly counting hybridization
line contractions scales exponentially with diagram order (and time).
In contrast, determinants can be computed at polynomial cost. However,
the dynamical sign problem, which always appears in non-equilibrium
Monte Carlo calculations, also incurs an exponential cost as a function
of time, so that both methods are exponential. In all applications
considered so far, the bold method converges at orders that are low
enough that the dynamical sign problem of the bare (or bold) expansion,
and not the exponential/polynomial scaling with diagram order, is
the limiting factor.

\begin{figure}
\includegraphics[width=8.6cm]{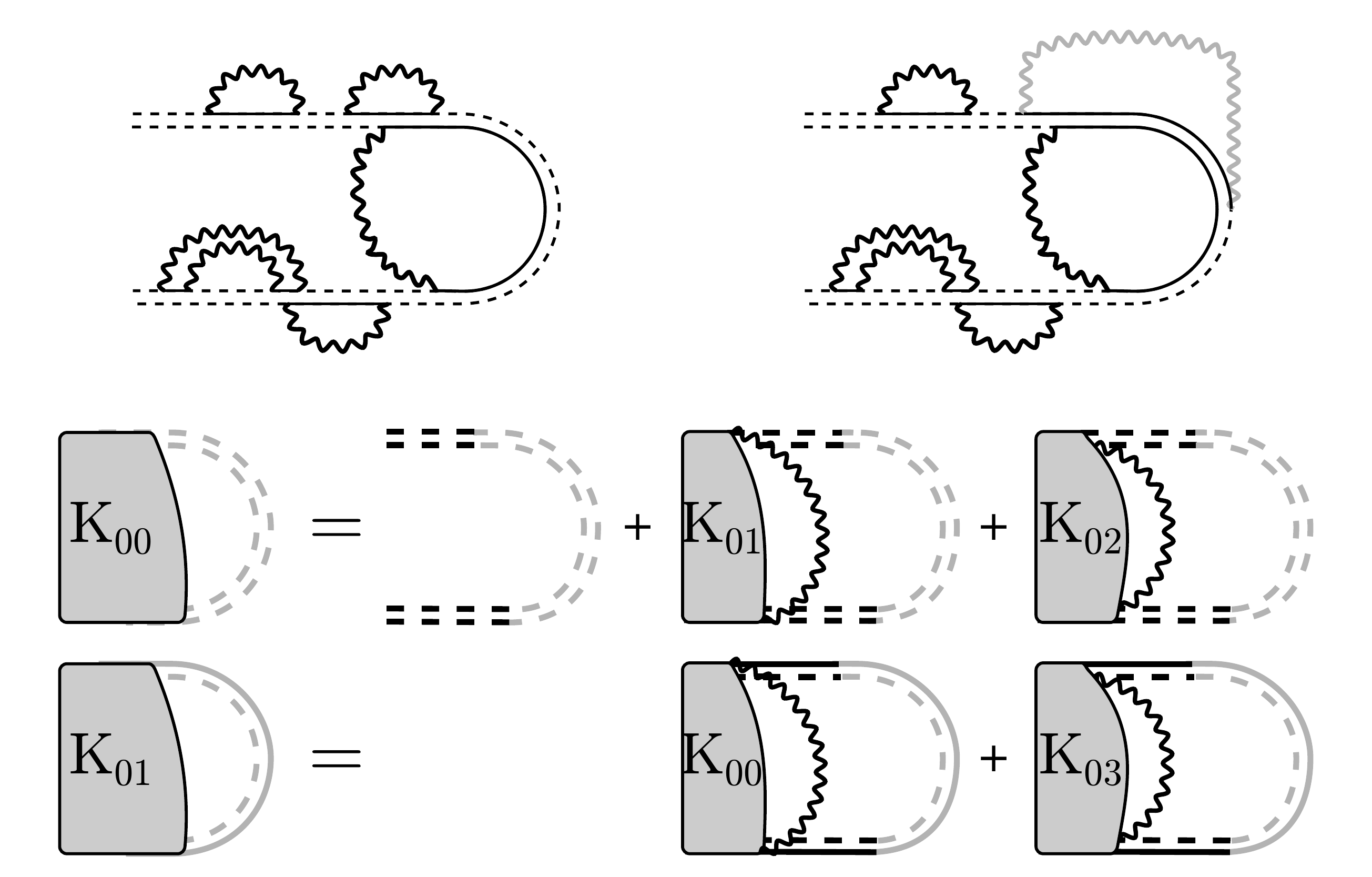}\caption{Top left panel: a diagram on the two-branch Keldysh contour. Note
the branch crossing hybridization line, which is not contained in
single-branch propagators. Top right panel: a diagram contributing
to the current or a two-time correlation function, with a special
hybridization line (grey) coupling to the final time. Lower panel:
two of the NCA vertex equations which take into account diagrams near
the beginning of the contour. The gray parts of the lines are not
part of the diagrams, but instead illustrate their location on the
Keldysh contour.\label{fig:keldysh}}
\end{figure}

So far, we have discussed only the propagators, which exist on a single
branch of the Keldysh contour. To obtain physical observables in real
time, one must take into account diagrams on both branches as well
as diagrams crossing branches. The structure of such inter-branch
diagrams is illustrated on the top left of Fig.~\ref{fig:keldysh}.
The example shows a contribution to the probability that the dot is
in the $\alpha=|\downarrow\rangle$ state at the final time at the
right tip of the contour, while beginning in state $|0\rangle$. Inter-branch
hybridization lines are a crucial part of the expansion, as they allow
changes to the state at the tip.

Generalizing the Monte Carlo process to the double contour poses no
difficulties greater than those entailed by the extra bookkeeping
(due to the fact that vertices on the two branches contribute opposite
signs, while time on the two branches flows in opposite directions).
However, the dressed propagators contain no information about inter-branch
diagrams. Just as one can use the bold line technique to renormalize
the propagators, it is possible and highly advantageous to obtain
partial summations of contour-crossing diagrams into vertex corrections.
An example of a set of self-consistent equations defining a vertex
correction is shown in the lower panel of Fig.~\ref{fig:keldysh}.
This particular vertex function, which we call the $K$ vertex, sums
both non-crossing intra-branch and inter-branch diagrams. It gives
all such contributions starting from some initial condition (marked
by the left index, 0 here) at the beginning of the contour and terminating
at some final condition (the right index) at a pair of propagation
times, one of which is on either branch. The rest of the contour (shown
in gray) can then be filled by any bold diagram not containing contributions
already summed within the vertex.

The two equations shown are part of a set of four coupled integral
equations which must be solved self-consistently, as they also depend
on $K_{02}$ and $K_{03}$. The final two equations have an analogous
structure. There are four such sets, one for each value of the initial
condition (see also Ref.~\onlinecite{gull_numerically_2011} and
Ref.~\onlinecite{cohen_numerically_2013}, which used a notation
that compacts all 16 $K$-vertex equations into one equation and also
includes the OCA contribution). Since the 16 elements of the vertex
are two time objects, they are stored in discretized form in large
$2D$ matrices. Currently a uniform discretization is used and the
memory requirements for the partial summation part of the method is
a bottleneck of the method. All integrals within the NCA vertices
and all but a few within the OCA take the form of a convolution (for
a Hamiltonian with no explicit time dependence), and can be performed
efficiently in Fourier space. The remaining integrals are amenable
to parallelization on clusters, and current implementations scale
linearly to over a thousand nodes on high-performance clusters.

Two important examples of observables which will be presented in the
results below are the electronic current into the different leads
and the two-time correlation function. Both observables are given
by diagrams which have a special line at the end of the Keldysh contour,
shown as the light gray line on the upper right corner of Fig.~\ref{fig:keldysh}.
This line is created by an operator $d_{\sigma}^{\dagger}$ at the
final measured time. To measure the current with spin $\sigma$ out
of certain leads, the hybridization function for the special line
should be replaced by the one defined by Eq.~\ref{eq:hybridization_function_definition},
with the sum over $k$ restricted to values within the chosen subset
of leads (for instance, to measure the ``left'' current one only
sums over $k$ values in the left lead). To measure two-time correlation
functions, the value of the hybridization line becomes unity (as it
describes free operators $d^{\dagger}\left(t\right)$ and $d\left(t^{\prime}\right)$
which do not come with a corresponding lead operator and coupling),
and the contribution is binned with respect to the location of the
operator at the other end of the line. With combinations of such lines
one can construct any single particle correlation function ($G^{<}$,
$G^{r}$ etc.).

As mentioned earlier, in practice all possible contractions of hybridization
lines from a diagram are taken into account simultaneously; this means
that the final expression for a contribution to (say) the current
from any given diagram is identical to the one introduced in Eq. (62-66)
of Ref.~\onlinecite{werner_diagrammatic_2009}, but with terms already
accounted for by the partial resummation excluded from the evaluation
of the determinant of Eq. (59) of that reference.

In using bold-line CTQMC within the context of reduced dynamics for
dot observables,\cite{cohen_memory_2011-1,cohen_numerically_2013}
the somewhat more complex $\Phi$ observables were implemented which
measure the current through one spin for a preselected occupation
of the other spin. It should be clear from the description above how
the generalization proceeds: one accepts only current diagrams where
the opposite spin meets the desired criterion. In fact, since both
possible criteria are needed for the reduced dynamics, all contributions
to the single-spin current can be sampled in each Monte Carlo trajectory,
as long as during the measurement only the appropriate observable
for each configuration is updated. A reduced dynamics treatment which
also allows for the computation of non-dot observables such as the
current would require a further addition of observables to the simulation.\cite{cohen_generalized_2013}

\section{Results\label{sec:Results}}

\begin{figure*}
\includegraphics[width=0.33\textwidth]{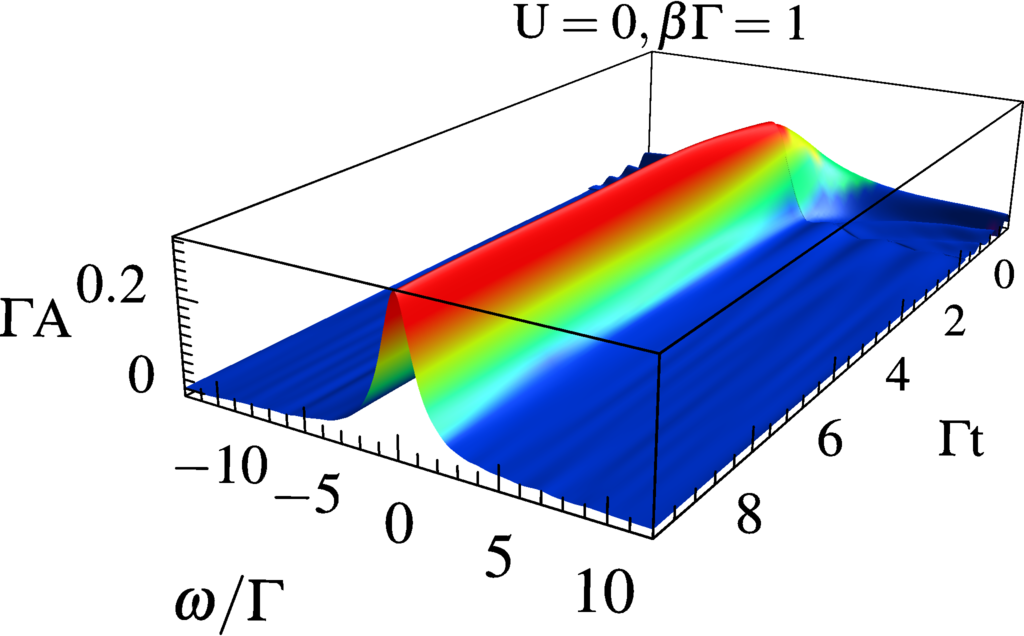}\includegraphics[width=0.33\textwidth]{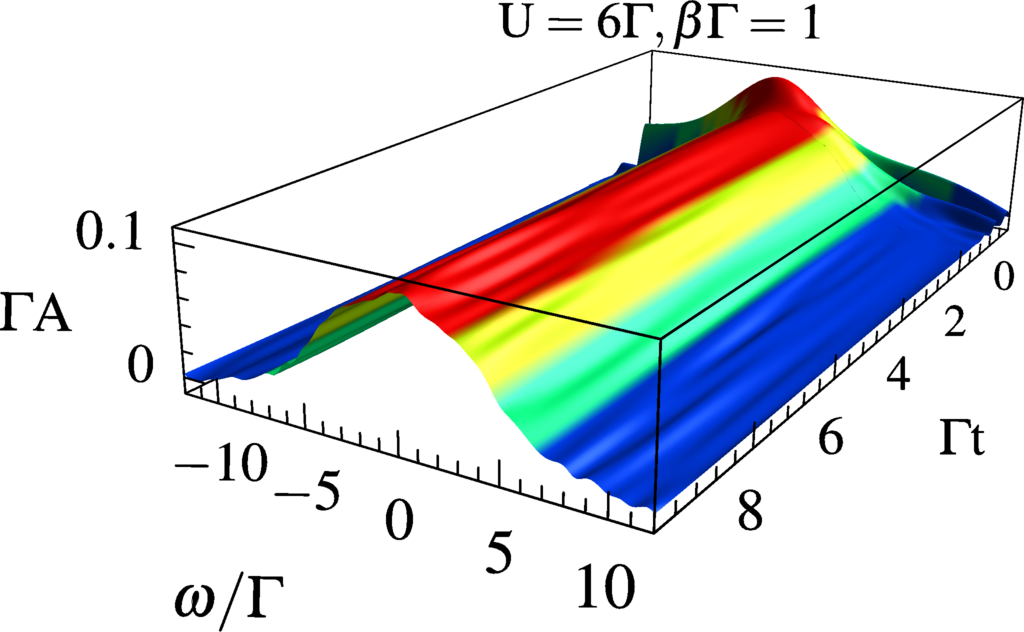}\includegraphics[width=0.33\textwidth]{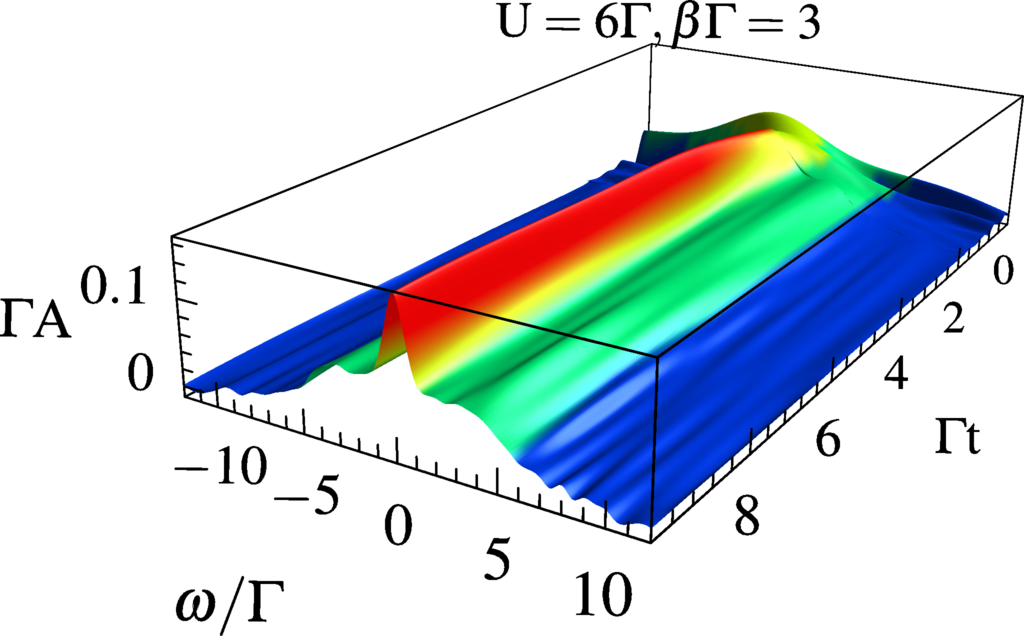}\caption{The dependence of the probed Spectral function $A\left(\omega\right)\equiv-\frac{1}{\pi}\Im\left\{ G^{R}\left(\omega\right)\right\} $
on time and frequency, as obtained from the auxiliary current method
for an initially decoupled quantum dot at $\Omega_{c}=10\Gamma$.
At $U=0$ and $\beta\Gamma=1$ (left), a simple Lorentzian shape develops.
At $U=6\Gamma$ and $\beta\Gamma=1$ (center), the interaction distorts
and widens the spectrum. At $U=6\Gamma$ and $\beta\Gamma=3$ (right),
the spectral profile typical to the Kondo problem develops, exhibiting
a central Kondo peak between two lower Hubbard peaks.\label{fig:A_t_three_cases}}
\end{figure*}

We now show results for the spectral function of the Anderson impurity
model, obtained using bold-line CTQMC and the auxiliary current method
and a direct measurement of the time-dependent Green's function. To
constrain the parameter space we restrict ourselves to the particle-hole
symmetric case $\varepsilon_{i}=-\frac{U}{2}$ and to a single flat
band with a soft cutoff $\Gamma\left(\omega\right)=\frac{\Gamma}{\left(1+e^{\nu\left(\omega-\Omega_{c}\right)}\right)\left(1+e^{-\nu\left(\omega+\Omega_{c}\right)}\right)}$.
We choose $\Gamma\nu=10$ and hold the lead chemical potential at
0, while varying dot interaction $U$, the lead temperature $\beta$,
and the band cutoff energy $\Omega_{c}$. All results are computed
with the real time bold-line CTQMC expansion\cite{gull_numerically_2011,cohen_numerically_2013}
built around the one-crossing approximation (OCA) \cite{eckstein_nonequilibrium_2010}
by measuring the current to an auxiliary reservoir defined by 
\begin{equation}
\Gamma_{A}\left(\omega_{A},\omega\right)=\frac{\eta\beta_{A}}{\sqrt{\pi}}e^{-\left[\beta_{A}\left(\omega-\omega_{A}\right)\right]^{2}},
\end{equation}
with $\Gamma\beta_{A}=10$ and $\eta=10^{-3}\Gamma$. The dot is initially
decoupled from the physical and auxiliary leads, and the coupling
is turned on at time zero, after which the system time-evolves according
to the full Hamiltonian, Eq.~\prettyref{eq:hamiltonian}.

Little is known about the properties of the spectral function as measured
from the auxiliary current, Eq.~\prettyref{eq:A_double_probe_scheme},
as a function of time; what we have shown rigorously is only is that
at steady state or equilibrium it must approach the spectral function
known from other methods. At intermediate timescales this quantity
contains information about quench dynamics (since the dot begins in
an out-of-equilibrium state where it is decoupled from the bath),
but the interpretation of this data must be performed with care, since
the physical dynamics are mixed to some degree with those of the auxiliary
lead. We therefore begin the discussion by exploring it in detail
for the three examples shown in Fig.~\ref{fig:A_t_three_cases}.
The $A\left(\omega\right)$ as measured by our virtual probes is plotted
as a function of time. Note that a constant-$t$ cut across the surface
at long times forms the steady state spectral function, which can
be read from the profile of the plots. With the initial ($t=0$) condition
we have chosen, the auxiliary current $I_{A}(t=0)$ is always zero.
In all three cases, the band cutoff energy of the leads $\Omega_{c}$
is set to $10\Gamma$; the band can therefore be considered essentially
flat over the range of frequencies displayed. The long time limit
corresponds to the equilibrium situation.

On the left, we display a noninteracting case. Within hybridization
expansion CTQMC, the noninteracting case is a stringent test for the
algorithm, as it expands about the atomic limit making this exactly
solvable limit a difficult case for our approach. First, the final
profile has a Lorentzian shape, as expected from a noninteracting
dot coupled to a flat lead. Second, the observed relaxation timescale
appears to be related to $\Gamma$ rather than the auxiliary parameters
$\eta$ and $\beta_{A}$. This is typical for charge-related properties
in the system, and suggests that we are in fact measuring physical
system properties and not properties related to our choice of auxiliary
lead. This is expected, since the auxiliary lead is coupled to the
system only weakly, so that probing the dot only involves its linear
response characteristics. 

\begin{figure}
\includegraphics[width=8.6cm]{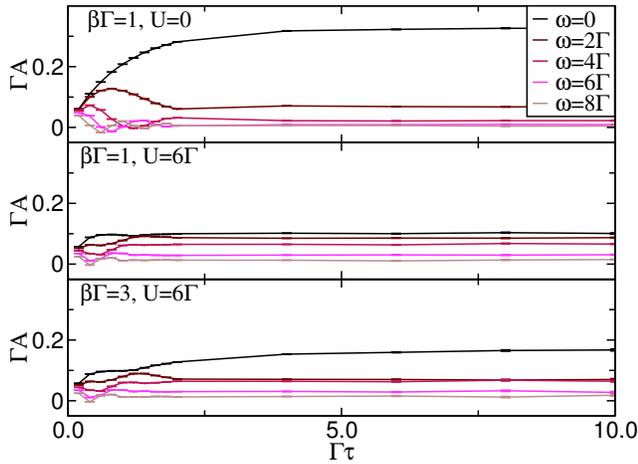}\caption{Several cuts through the data of Fig.~\ref{fig:A_t_three_cases}
are shown at fixed $\omega$, making it easier to observe both the
convergence and the behavior at short times.\label{fig:A_t_cuts}}
\end{figure}

The behavior at both short and long timescales can alternatively be
visualized by taking cuts through this data at planes of fixed $\omega$,
as shown in the top panel of Fig.~\ref{fig:A_t_cuts}. Here, we see
that relaxation to equilibrium occurs somewhat more slowly for energies
near the Fermi level. At short timescales the data contains information
about the nonequilibrium evolution of the system in a quench situation,
but this is mixed to some degree with the relaxation properties of
the auxiliary lead. Exploring the implications of this in detail is
beyond the scope of the current work.

In the middle panel of Fig.~\ref{fig:A_t_three_cases}, a strong
interaction has been turned on, while the temperature is kept rather
high. At long times, excitations spread throughout the band, but not
far beyond $\Omega_{c}$. In addition, noise and oscillations in both
time and frequency appear, indicating that convergence to the steady
state is much slower. However, as before, the time evolution of the
spectral function is mostly converged after a time on the order of
$\frac{1}{\Gamma}$ (see also the corresponding middle panel of Fig.~\ref{fig:A_t_cuts}).

On the right side of Fig.~\ref{fig:A_t_three_cases} the temperature
is lowered while the interaction is reduced. The formula $k_{B}T_{K}=U\sqrt{\frac{\Gamma}{2U}}e^{-\frac{\pi U}{8\Gamma}+\frac{\pi\Gamma}{2U}}$
is commonly used to estimate the Kondo temperature\cite{hewson_kondo_1993}
and gives $\Gamma\beta_{K}\simeq4.7$, yet even at $\Gamma\beta_{K}\simeq3$
the predicted spectrum clearly begins to display the characteristics
of Kondo physics, and one can observe the formation of a central peak
at the chemical potential and an indication of two side bands. Interestingly,
the time scale over which the general profile develops does not appear
to be significantly modified, but long-time oscillations which have
not fully attenuated by our final simulation time appear. This suggests
that obtaining numerical convergence for this particular parameter
set requires longer propagation times than required in the strongly
interacting regime. On the other hand, it also suggests that obtaining
a basic estimate of the spectral function only requires comparatively
short propagation times (see also the corresponding lower panel of
Fig.~\ref{fig:A_t_cuts}).

\begin{figure}
\includegraphics[width=8.6cm]{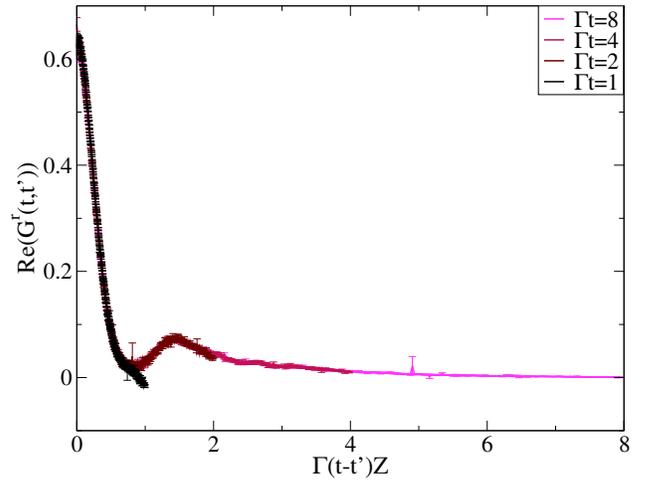}\caption{The real part of the retarded two-time correlation function $G^{r}\left(t,t^{\prime}\right)$
as a function of the time difference $t-t^{\prime}$, shown for for
several values of the final time $t$. The parameters chosen are $U=6\Gamma$,
$\beta\Gamma=3$ and $\Omega_{C}=10\Gamma$.\label{fig:G_two_times_convergence}}
\end{figure}

As explained in \prettyref{sec:Bold-line-Monte-Carlo}, within bold-CTQMC
two-time correlation functions can also be obtained directly for a
given $t$ and $t^{\prime}$. The result of one such evaluation for
$G^{r}\left(t,t^{\prime}\right)$ is displayed in Fig.~\ref{fig:G_two_times_convergence}.
At equilibrium $G^{r}$ must become a function of the difference $t-t^{\prime}$
between its two time parameters, but for any finite $t$ it exists
only for $t^{\prime}<t$. We expect that for large enough $t$, the
correlation function measured as a function of $t-t^{\prime}$ should
converge to the equilibrium value; Fig.~\ref{fig:G_two_times_convergence}
illustrates that while this clearly is not the case for $\Gamma t\lesssim2$,
convergence occurs rather quickly. For the parameters shown, at $\Gamma t\gtrsim4$
one obtains reliable estimates of two-time correlation functions in
equilibrium for any $t-t^{\prime}<t$.

Once the equilibrium correlation function $G^{r}\left(t-t^{\prime}\right)$
is obtained in the time domain, a simple Fourier transform takes us
to the frequency domain, where the spectral function may be obtained
using Eq.~\ref{eq:A_from_Gs}. In practice, since $t-t^{\prime}$
is limited in range by the maximum $t$ reachable, one must also converge
the result of the Fourier transform in $t$ at all frequencies. This
proves to be difficult, and we will show that when one is interested
in frequency domain properties the auxiliary current method offers
more accurate results. However, for certain nonequilibrium problems,
it is often the time-domain function itself that proves to be of interest.\cite{cohen_greens_2013,eckstein_thermalization_2009}

\begin{figure}
\includegraphics[width=8.6cm]{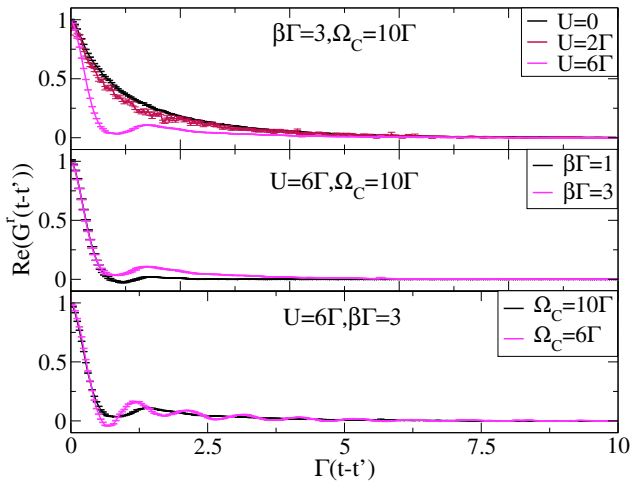}\caption{The real part of the retarded two-time correlation function $G^{r}\left(t,t^{\prime}\right)$
as a function of the time difference $t^{\prime}-t$, for a final
time $t$ of $10/\Gamma$ . The dependency on different parameters
is illustrated by taking different values of $U$ (top panel), $\beta$
(middle panel) and $\Omega_{C}$ (lower panel) with the other parameters
fixed.\label{fig:G_two_times_different_parameters}}
\end{figure}

The dependence of the real part of the retarded equilibrium Green's
function on the physical parameters of the system is explored in Fig.~\ref{fig:G_two_times_different_parameters}.
The top panel shows that while the effect of weak interaction is relatively
mild, strong interaction induces qualitative changes in the structure
of the correlations. The central panel shows that these effects are
mitigated but not destroyed by higher temperatures. Finally, the bottom
panel shows how a reduction in the bandwidth of the bath results in
the introduction of high-frequency oscillations into the correlation
function.

\begin{figure}
\includegraphics[width=8.6cm]{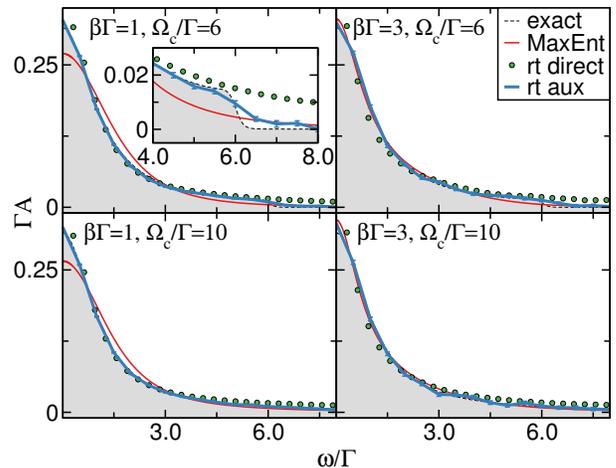}\caption{The spectral function $A\left(\omega\right)$ in the non-interacting
limit $U=0$ is shown for several combinations of the inverse temperature
$\beta\Gamma$ and band width $\Omega_{c}/\Gamma$. The exact result
is shaded in gray for comparison to the imaginary time data analytically
continued with MaxEnt using a flat model (solid red), the result obtained
from directly Fourier transforming the real time bold-line CTQMC correlation
function (green circles), and the real time bold-line CTQMC auxiliary
current data (thick blue line with error bars). Both real time results
are obtained by propagating a decoupled initial state to $\Gamma t=10$.
The inset in the top left panel zooms in on the region near the band
edge.\label{fig:noninteracting}}

\end{figure}

To validate our results, Fig.~\ref{fig:noninteracting} compares
the spectral function, as obtained with real time bold-line CTQMC
from both the auxiliary current method and the direct Fourier transformation
of the time-domain correlation function, to the exact spectrum in
the noninteracting case, shown as the gray shaded region terminated
by a dashed black line. We note that this is a particularly difficult
limit for bold-line CTQMC based on the OCA, as OCA performs poorly
for small values of $U$; to decrease the amount of computer time
needed, we have limited the maximum order of bold diagrams to $6$,
despite the fact that higher order contributions are contributing
on the order of a few percent. As a basis for assessing the accuracy
of the proposed method compared to existing implementations, we also
show a result obtained from a simulation of the same model using the
imaginary time hybridization Monte Carlo method,\cite{werner_continuous-time_2006},
analytically continued using the Maximum Entropy\cite{jarrell_bayesian_1996}
algorithm (MaxEnt) with an unbiased flat default model.

Fig.~\ref{fig:noninteracting} clearly shows that none of the numerical
methods perfectly reconstruct the exact spectral function. The real
time auxiliary current method (blue line with error bars denoting
statistical, but not finite time or order truncation errors) is the
only one to perform consistently well at all parameters and frequencies
shown. However, some systematic errors remain, in particular at the
band edges and near the chemical potential. The Fourier-transformed
real time correlation function (green circles) is not as accurate,
and in particular displays a high-frequency tail caused by numerical
noise (the auxiliary lead formulation circumvents this by the finite
width of the auxiliary lead's coupling density. However, this always
induces an averaging over a finite frequency window).

The MaxEnt result computed with a flat model (solid red line) in Fig.~\ref{fig:noninteracting}
appears to behave relatively well only at low temperature and large
bandwidth, as seen in the bottom right panel. As we raise the temperature---going
from the bottom right to the bottom left panel---MaxEnt fails, in
particular at frequencies near the resonance. The peak is at odds
with the bias towards flatness present in our default model, and causes
a noticeable deviation from the correct result as we go from bottom
right to top right panel. While the MaxEnt results are incompatible
with the real time results, the back-continuation of the real-time
results to the imaginary axis (where imaginary time Monte Carlo results
are numerically exact) agrees perfectly with the imaginary time results,
and the back-continuation of the continued MaxEnt results also agrees
perfectly with the imaginary time results. The fact that two completely
incompatible real-frequency spectra can have the same Matsubara frequency
representation (within Monte Carlo errors on the imaginary axis) is
well known and caused by the large number of very small eigenvalues
of the analytic continuation kernel. Continuing further to the top
left panel, when both temperature is low and bandwidth is high, MaxEnt
results are unreliable both at high and at low frequencies (while,
again, imaginary frequency Green's functions agree within errors).
As seen in the inset, no signature of high energy features, e.g. the
band edge, is discernible. This also remains true at lower temperatures.

A few caveats regarding the comparison with MaxEnt are in order. First,
analytic continuation is an ill-posed problem, and the analytically
continued result depends strongly on the algorithm and parameters
used. In the case of MaxEnt, it also depends on the model. As the
algorithm is based on a penalty for deviations from the default model,
a correct guess for the noninteracting system would have lead to the
exact result. We therefore stress here that additional information
(e.g. from low-order perturbation theory or high-temperature simulations)
may yield substantially better results by taking advantage of additional
knowledge; it may however also be a source of bias. In the context
of real time method, the ill-posed continuation problem can be avoided
and the sources of ambiguity introduced when continuing results to
the real axis are absent from the start. 

\begin{figure}
\includegraphics[width=8.6cm]{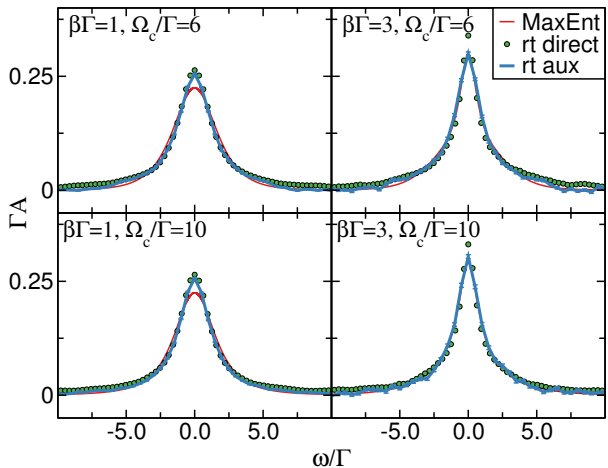}\caption{The spectral function $A\left(\omega\right)$ in the weakly interacting
limit $U=2\Gamma$ is shown for several combinations of the inverse
temperature $\beta\Gamma$ and band width $\Omega_{c}/\Gamma$. The
imaginary time data analytically continued with MaxEnt using a flat
model (solid red) is shown along with the result obtained from directly
Fourier transforming the real time bold-line CTQMC correlation function
(green circles), and the real time bold-line CTQMC auxiliary current
data (thick blue line with error bars). Both real time results are
obtained by propagating a decoupled initial state to $\Gamma t=10$.\label{fig:weakly interacting}}
\end{figure}

We now turn to the exploration of the results of the direct measurement
of the spectral function and of the auxiliary current method in combination
with bold-line CTQMC in interacting systems. In Fig.~\ref{fig:weakly interacting}
we increase the interaction strength to $U=2\Gamma$. The two real
time methods are in far better agreement in this case, but deviations
at high frequency are visible. Once again, only the real time auxiliary
current correctly captures the existence of a band edge, and the MaxEnt
method is unable to capture the results both for low band width and
high temperature.

The nonzero interaction makes convergence with the order of bold diagrams
much easier to obtain, as the quality of the OCA approximation improves
and the number of hybridization events decreases. Still, these weakly
interacting parameters remain beyond the scope where NCA or OCA are
accurate. In fact, even a perturbative weak coupling treatment would
be sufficient in this regime. The main result of this plot is therefore
that it highlights the ability of OCA based bold-line CTQMC to treat
interacting problems substantially beyond the range where OCA itself---without
Monte Carlo---is accurate.

\begin{figure}
\includegraphics[width=8.6cm]{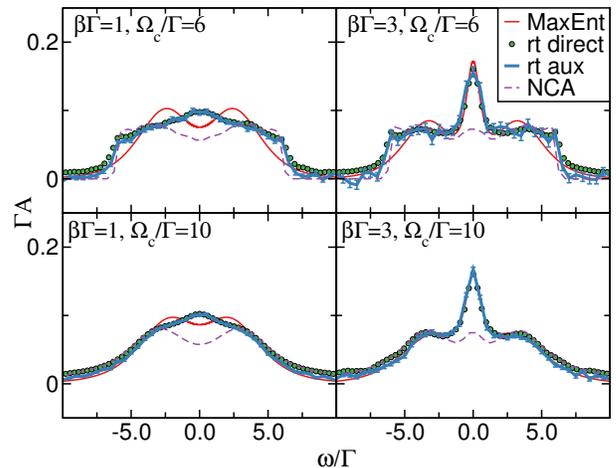}\caption{The spectral function $A\left(\omega\right)$ in the strongly interacting
limit $U=6\Gamma$ is shown for several combinations of the inverse
temperature $\beta\Gamma$ and band width $\Omega_{c}/\Gamma$. The
imaginary time data analytically continued with MaxEnt using a flat
model (solid red) is shown along with the result obtained from directly
Fourier transforming the real time bold-line CTQMC correlation function
(green circles), and the real time bold-line CTQMC auxiliary current
data (thick blue line with error bars). Both real time results are
obtained by propagating a decoupled initial state to $\Gamma t=10$.
Real time NCA data is also shown for comparison (dashed purple line).
\label{fig:strongly-interacting}}
\end{figure}

Next, we continue to a strongly interacting case, Fig.~\ref{fig:strongly-interacting},
with $U=6\Gamma$. Here $\frac{U}{\Gamma}$ is large enough that NCA
can be a useful approximation, and we include it for comparison. As
before, the two real time methods give consistent results at small
to intermediate frequencies. At large bandwidth and low temperatures
(lower right), all Monte Carlo methods, including imaginary time CTQMC,
yield similar results except at high frequencies, while the NCA underestimates
the height of the central resonance. However, when either the bandwidth
or the temperature is varied, a dramatic change occurs: lowering the
bandwidth (upper right) results in the development of a high and sharp
band edge, which the MaxEnt method misses completely. In this respect,
MaxEnt actually performs even worse than NCA (though results are clearly
much better at low frequencies). At higher temperature (lower left
of Fig.~\ref{fig:strongly-interacting}), both MaxEnt and NCA underestimate
the central resonance, resulting in a dip where a peak should be and
in a qualitatively different spectral shape at low frequencies. When
both temperature and bandwidth are varied simultaneously, NCA continues
to suffer from the same problem, while MaxEnt suffers from both simultaneously,
resulting in a qualitatively incorrect spectral function at all frequencies.

We note that previous authors have found that at least for some parameters,
a perturbative expansion including select diagrams beyond the OCA
results in substantially better approximations.\cite{eckstein_nonequilibrium_2010}
Using these higher order expansions as a starting point within bold-line
CTQMC is expected to improve its convergence properties further, making
it potentially easier to obtain reliable, unbiased data independent
of the underlying approximation.

\section{Summary and conclusions\label{sec:Summary-and-conclusions}}

To conclude, we have proposed and implemented two methods of obtaining
Green's functions from real-time bold-line CTQMC. The first one directly
evaluates two-time correlation functions, while the second one is
based on a newly proposed double-probe scheme for the computation
of an auxiliary current. The auxiliary current method is general enough
to be used with other numerical methods, and is especially helpful
when two-time correlation functions are impossible or more expensive
to acquire, as is the case in formalisms relying on reduced dynamics.
We compared these two methods to each other, to analytically continued
data and to data obtained within the non-crossing approximation. In
general the real time methods give more accurate results and the auxiliary
current method in particular even resolves high-energy features such
as hard band edges. Numerically exact convergence accounting for all
sources of errors remains technically difficult to achieve. The MaxEnt
method (in the particular variation we chose to test) does not reconstruct
high-frequency features. At low temperature it performs very well
at low frequencies, but at higher temperatures it fails. We demonstrate
that for certain choices of model parameters, which are neither special
nor extreme, both MaxEnt and NCA produce qualitatively misleading
results.

Real time methods remain substantially more expensive than imaginary
time methods in terms of the computer time needed to accomplish similar
tasks, and will in all probability remain so. As DMFT continues to
evolve towards being a general-purpose tool for material science,
it is plausible to assume that better scaling methods capable of treating
larger impurities with more orbitals per site will become standard,
even for real-time dynamics. While in this paper we have considered
only expansions about the NCA and OCA for a single-site Anderson impurity
model, the ideas and formalism are general: they can also be used
for other formal resummations, or for expansions about more complicated
models such as those arising in cluster extensions of dynamical mean
field theory. Even now, real time methods offer (aside from the ability
to address nonequilibrium situations, which was discussed elsewhere\cite{cohen_greens_2013})
an alternative and controlled route to access spectral quantities
at high frequency that is not available with other tools, and will
find its use in particular in real-time DMFT.
\begin{acknowledgments}
The authors would like to thank Rainer Härtle for his useful input
on many occasions. GC is grateful to the Yad Hanadiv--Rothschild Foundation
for the award of a Rothschild Postdoctoral Fellowship. GC and EG acknowledge
TG-DMR120085 and TG-DMR130036 for computer time. DRR acknowledges
NSF CHE-1213247, AJM acknowledges NSF DMR 1006282, and EG acknowledges
DOE ER 46932. This research used resources of the National Energy
Research Scientific Computing Center, which is supported by the Office
of Science of the U.S. Department of Energy under Contract No. DE-AC02-05CH11231.
Our implementations were based on the ALPS \cite{bauer_alps_2011}
libraries.
\end{acknowledgments}
\bibliographystyle{apsrev4-1}
\bibliography{Library}

\end{document}